\documentclass[a4paper,11pt]{article}
\usepackage{pos}
\usepackage{float}
\usepackage{natbib}
\usepackage{graphicx}
\usepackage{subcaption}
\usepackage{enumitem}
\usepackage{gensymb}

\newlength{\bibitemsep}
\newlength{\bibparskip}
\let\oldthebibliography\thebibliography
\renewcommand\thebibliography[1]{%
  \oldthebibliography{#1}%
  \setlength{\parskip}{\bibitemsep}%
  \setlength{\itemsep}{\bibparskip}%
}

\title{Simulating Geomagnetic Effects on Muons in Extensive Air Showers for the EUSO-SPB2 Mission}
 \ShortTitle{Geomagnetic Effects of Muons in EAS}

\author*[a]{Duncan Fuehne}
\author[a]{Tobias Heibges}

\affiliation[a]{Colorado School of Mines, Physics Department\\
  1500 Illinois St., Golden, Co, USA}

\onbehalf{for the JEM-EUSO Collaboration\\[-1mm]{\normalsize \normalfont
(a complete list of authors can be found at the end of the proceedings)}}

\emailAdd{dfuehne@mines.edu}
\emailAdd{theibges@mines.edu}

\abstract{The Extreme Universe Space Observatory on a Super Pressure Balloon II (EUSO-SPB2) measured extensive air showers (EASs) from upward-going High Energy Cosmic Rays by flying a Cherenkov Telescope (CT) at 33 km altitude. The telescope could be tilted just above the Earth’s limb, 5.8° below horizontal, and 650 km away as viewed from the balloon. This configuration enables the detection of EASs that develop over a longer path length than downward-going showers. The lifetime of 100 GeV muons, as an example, corresponds to a path length of 620 km in Earth's upper atmosphere, where there is a decreased amount of energy lost due to atmospheric interactions (only $\approx 40 \:\text{MeV/km}$ lost at 15 km altitude). In this configuration, muons can travel hundreds of kilometers while bending in Earth's geomagnetic field before they decay. These effects cause EASs to be more dense with muons at larger shower depths compared to the shower at $X_{\text{max}}$, a result known as the \textit{muon tail}. The objective of this simulation is to understand whether the CT on EUSO-SPB2 could measure the Cherenkov signal produced by the \textit{muon tail} and observe the effects of the muons deflecting in Earth's geomagnetic field. We found that the timing and angular distributions of the Cherenkov signal allow the muon component to be separated from the main Cherenkov signal and we identified quantifiable effects of the muons deflecting in the geomagnetic field. However, at this time we are unable to simulate enough events to analyze the distribution of photons arriving in an area the size of the aperture of the CT. Thus, we cannot make conclusions about whether these effects can be seen by EUSO-SPB2.
}

\ConferenceLogo{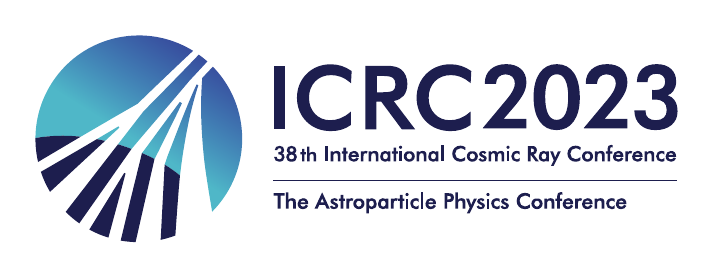}

\FullConference{%
38th International Cosmic Ray Conference (ICRC2023)\\
  26 July - 3 August, 2023\\
  Nagoya, Japan}

\begin{document}
\maketitle

\section{Introduction \& Simulation Overview}
\indent The Extreme Universe Space Observatory on a Super Pressure Balloon II (EUSO-SPB2) was designed and built to observe extensive air showers (EASs) created by High Energy and Ultra-High Energy Cosmic Rays (HECR \& UHECR) from Earth's upper atmosphere. After a successful launch from Wanaka, New Zealand, EUSO-SPB2 reached the planned float altitude of 33 km. Unfortunately, a leak in the balloon forced an early flight termination after two nights of collecting data. The payload featured two optical astroparticle telescopes. A Fluorescence telescope was pointed downwards to observe fluorescence light from EASs. A Cherenkov Telescope (CT) was pointed either just above or below the Earth's Limb, 5.8° below horizontal and 650 km away at this altitude. When pointed above the limb, the CT recorded some EAS candidates consistent with the expected optical signatures of PeV energy EASs in Earth's upper atmosphere.

\begin{figure}[h]
\begin{centering}
\begin{subfigure}{0.6\textwidth}

    \includegraphics[width = \textwidth,trim={0 0 0 0.5cm},clip]{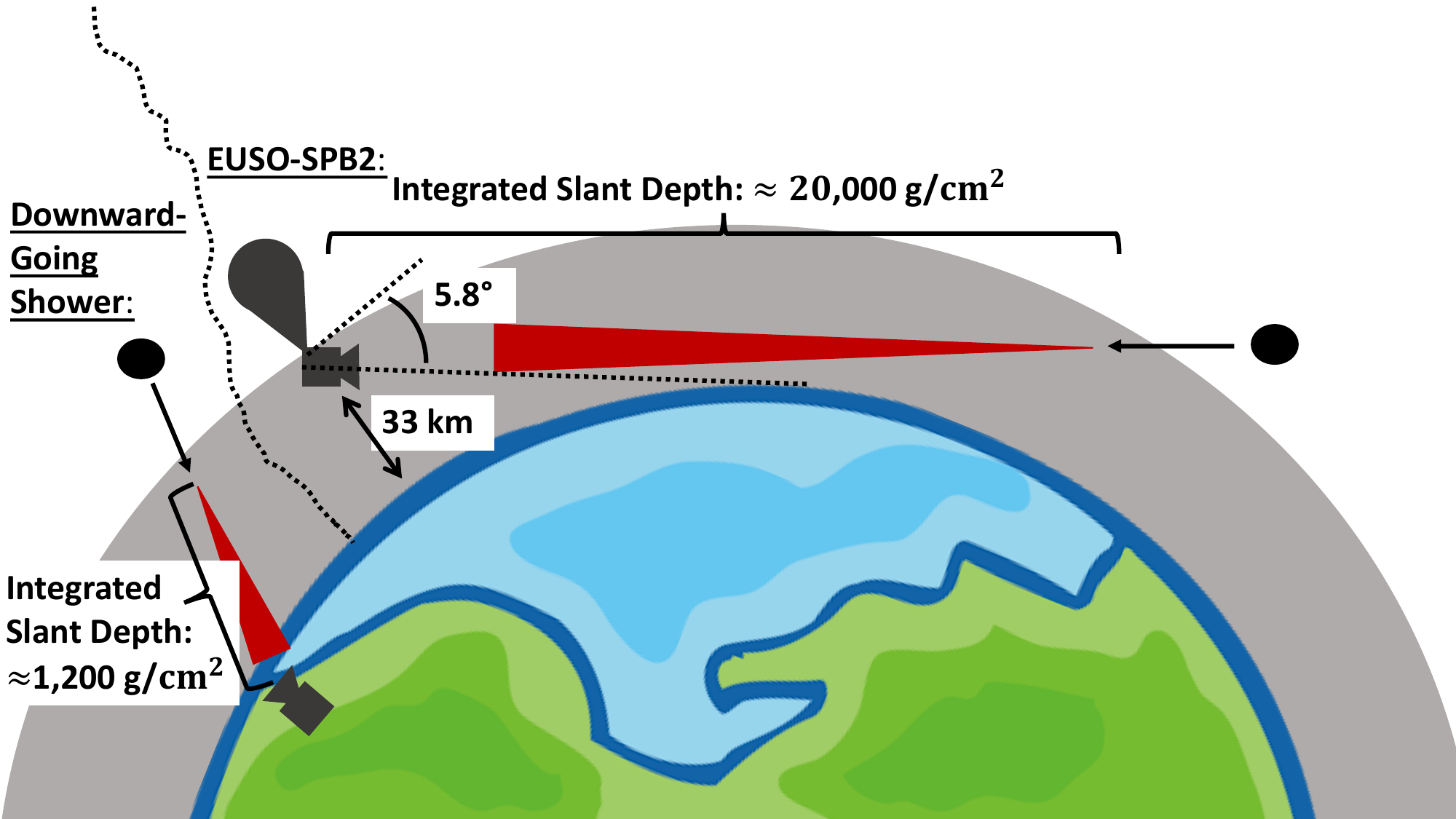}
    \caption{}
\end{subfigure}
\hfill
\begin{subfigure}{0.35\textwidth}
    \includegraphics[width = \textwidth]{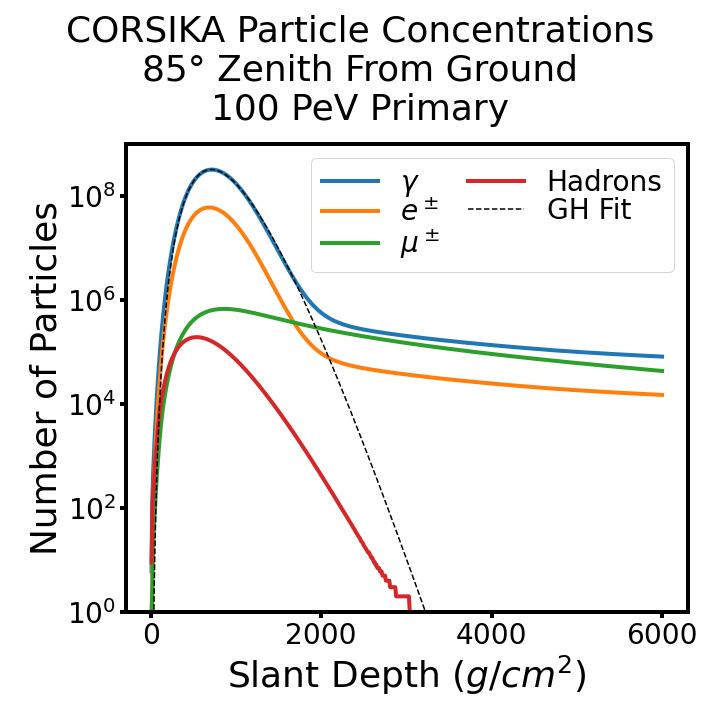}
    \caption{}
\end{subfigure}
\end{centering}
\caption{\textbf{(a)} Differences in geometry for EUSO-SPB2 vs. Ground-based detectors. EUSO-SPB2 was sensitive to EASs that traversed a much longer slant depth than downward-going EASs. The slant depth of 20,000 $\text{g/cm}^2$ represents an EAS arriving at the balloon from 0.5$\degree$ above Earth's limb \cite{Cummings_2021}. \textbf{(b)} Particle numbers for a 100 PeV air shower modeled in CORSIKA. The shower started at the ground with a zenith angle of 85° (5° above horizontal). Tables for Figure provided by Fred Garcia of the nuSpaceSim collaboration  \cite{nuSpaceSim}.}
\label{fig:geo}
\end{figure}

As seen in Figure \ref{fig:geo}, EASs measured in this configuration traverse a much longer distance than downward EASs. For instance, when the CT was pointed 0.5$\degree$ above Earth's limb, it was sensitive to EASs that traversed $20,000 \: \text{g}/\text{cm}^2$ of atmosphere, whereas an air shower traveling downwards towards Earth would have an integrated slant depth of just over $1,000 \: \text{g}/\text{cm}^2$ \cite{Cummings_2021}. Most of the EAS's path occurs in Earth's upper atmosphere. At 15 km altitude and above, atmospheric interactions will only decrease a muon's energy by less than 40 MeV/km. Furthermore, the lifetime of 100 GeV muons corresponds to a path length of 620 km before they decay \cite{muondecay}.  Both of these factors allow muons to travel hundreds of kilometers inside the EAS, giving rise to a long \textit{muon tail}. Figure \ref{fig:geo} highlights this \textit{muon tail}, where the number of muons surpasses the number of electrons and approaches the number of gammas in the shower after a slant depth of about $2,000\: \text{g}/\text{cm}^2$. This feature is not described by the Gaisser-Hillas (GH) \cite{GassierHillas} parameterization. Since muons endure to longer slant depths (compared to other particles), their deflections in Earth's magnetic field should be more noticeable. The objective of this simulation is to model the \textit{muon tail} and determine if its Cherenkov signal could be observed by the CT. We then attempt to understand which features in this Cherenkov signal exemplify the effects of the muons deflecting in Earth's geomagnetic field.

\section{Simulation Methodology}
\indent The first step in the simulation is to choose the geometry for simulating the EAS, including the altitude and location (latitude and longitude) of the telescope. We then define the optical axis of the telescope by choosing the angle above Earth's limb and azimuthal angle (with respect to magnetic north). Assuming the shower axis of the EAS occurs along the optical axis of the telescope, we model a given number of muons with creation locations randomly selected according to the GH distribution \cite{GassierHillas}. We can use the GH distribution for creating the muons because it accurately models the longitudinal distribution of hadrons, including charged pions that decay to produce the muons in the tail. We define the muon energy distribution so that the number of muons ($N_{\mu}$) depends on their kinetic energy ($K_\mu$) via a power law with a spectral index of -2: $\left( N_{\mu} \propto (10\:\text{GeV} + K_{\mu})^{-2} \right)$. The 10 GeV scales the distribution centroid to a realistic magnitude. We also do not include a lateral momentum distribution as a first-order estimate (all muons travel along the shower axis). More future analysis with CORSIKA will be done to refine these two input distributions.

\subsection{Propagation of Muons Through Atmosphere}
\indent Once the muons are initialized with positions, energies, and momentum directions, the simulation individually steps each muon through Earth's geomagnetic field and atmosphere, including the four physical processes described below.

\textbf{Magnetic field deflection} -- The particle interacts with the earth's geomagnetic field through Equation \ref{eqn:magfield} \cite{BrauCharlesA.2004Mpic}. $u^\mu$ and $x^\mu$ are the muon's instantaneous four-velocity and four-position, indexed by $\mu$. $s$ is the proper time of the particle, $q$ is the particle's charge and $m$ is the particle's mass. The electromagnetic field tensor (indexed by $\mu$ and $\nu$), ${F_{\nu}}^\mu$ contains the components of Earth's geomagnetic field at every step along the particle's trajectory. 
\begin{equation} \label {eqn:magfield}
    \frac{d u^\mu}{ds} = \frac{q}{m} {F_\nu}^\mu u^\nu \quad \& \quad \frac{dx^\mu}{ds} = u^\mu
\end{equation}
These coupled ordinary differential equations are solved numerically in the lab frame using a 4th Order Runge-Kutta technique at each step of duration 100 ns proper time \cite{ode}. The components of the geomagnetic field are calculated using the International Geomagnetic Reference Field (IGRF) parameterization \cite{IGRF}.

\textbf{Energy loss} -- R. L. Workman of the Particle Data Group (PDG) parameterizes muon energy loss in the air as a function of its energy, accounting for ionization losses as well as expected values for losses due to bremsstrahlung, direct pair production, and photo-nuclear interactions    \cite{PPM}. We interpolate this table to determine the total energy loss the muon experiences at each step. Currently, we are working on implementing Monte Carlo methods for the stochastic energy loss effects.

\textbf{Cherenkov light production} -- Muons produce Cherenkov light throughout their trajectory. In each step of length $dx$, the number of photons ($N_\gamma$) of wavelength $\lambda$ emitted at angle $\theta_C$ takes the form of Equation \ref{eqn:cherprod} where $\alpha$ is the fine structure constant, $z$ is the charge number of the muon, $\beta$ is the fraction of the speed of light at which the muon is traveling, and $n(\lambda)$ is the refractive index of the material through which the particle is traveling \cite{PPM}. 
\begin{equation} \label{eqn:cherprod}
    \frac{d^2N_\gamma}{d\lambda dx} = \frac{2 \pi \alpha z^2}{\lambda^2} \left(1 - \frac{1}{\beta^2 n(\lambda)^2} \right) \quad \& \quad \cos(\theta_C) = \frac{1}{n(\lambda) \beta}
\end{equation}
We determine the total number of Cherenkov photons produced in each step by analytically integrating this expression with respect to wavelength from 200 nm to 1000 nm using an explicit form of $n(\lambda)$ \cite{index1, index2}. The form of $n(\lambda)$ given in \cite{index2} makes evaluating this integral unfeasible, so we take advantage of the fact that $n(\lambda) - 1 \ll 1$ to approximate $n(\lambda)^2$ in Equation \ref{eqn:cherprod} with a binomial approximation. Once the integral is evaluated, we have a functional form for the total number of Cherenkov photons produced between 200 nm and 1000 nm. Since there are different values of $\theta_C$ for different wavelengths, the maximum value is returned by the function. Later on, the simulation creates an angular distribution using smaller angles to account for its wavelength dependence.

\textbf{Muon decay} -- During each step of period, $\Delta t$, the muons have a non-zero probability to decay with a lifetime of $\tau = 2.197 \:\mu \text{s}$ \cite{muondecay} following an exponential decay law. Because we measure the time interval in the lab frame, the lifetime is multiplied by the Lorentz factor, $\gamma$, making the decay energy-dependent \cite{BrauCharlesA.2004Mpic}. We use only the dominant muon decay branch, $\mu^\pm \to e^\pm + \nu_\mu + \nu_e$ in this simulation (one neutrino is an antineutrino depending on the charge of the muon).

Muon propagation and Cherenkov light production occur until either the muon passes the plane of the detector, the muon loses all of its energy, or it decays. In all three cases, the total Cherenkov production is recorded but in the first two cases, the process ends. However, if the muon decays then the electron/positron (in the rest of the paper, "electron" refers to both electrons and positrons) goes into the next step of the simulation. The total energy of that electron in the center-of-mass frame of the muon is determined by the electron energy-dependent decay rate of muons, which was experimentally measured in \cite{muondecay}. We randomly select the momentum of the electron to point anywhere in three-dimensional space in the muon rest frame. Then the electron energy and momentum are Lorentz-boosted into the lab (balloon) frame. \cite{BrauCharlesA.2004Mpic}.

\subsection{Propagation of Electrons Through Atmosphere}
\indent The total Cherenkov light production from electrons is estimated as follows. However, we currently only track the single electrons created from muon decay and have not yet implemented any methods of accounting for the electromagnetic shower of particles created from these electrons. Nevertheless, we have a first-order estimate of the Cherenkov production and its distributions starting from first principles. The simulation propagates the electrons through the atmosphere in a similar way to the muons until the electrons run out of energy. The following are the physical effects considered during each step of electron propagation.

\textbf{Magnetic field deflection} -- Calculations for the geomagnetic field effects on electrons use the same methodology as with the muons.

\textbf{Multiple scattering} -- Electrons have a higher chance of multiple scattering in the atmosphere than muons. We define the average scattering angle, $\theta_0$ in terms of the momentum ($p$), speed ($\beta c$), and charge number ($z$) of the electron \cite{PPM}. The scattering angle is also dependent on the distance traveled in each step ($x$) per radiation length of the atmosphere ($X_0$).  
\begin{equation} \label{eqn:scatter}
    \theta_0 = \frac{13.6 \:\text{MeV}}{p\beta c} z \sqrt{\frac{x}{X_0}} \left[1 + 0.038 ln\left(\frac{xz^2}{2X_0 \beta^2}\right)\right]
\end{equation}
In each step, two independent, Gaussian random variables are created which define the angle and position offset due to multiple scattering. The angular offset is centered around $\theta_0$ and the position offset is centered around the lateral distance the particle would travel if it were deflected by $\theta_0$ halfway through the step \cite{PPM}.

\textbf{Energy loss} -- The two mechanisms that contribute most to electron energy loss in the atmosphere are bremsstrahlung and ionization. Given any electron with mass $m_e$ and energy $E$, Equation \ref{eqn:elecionization} defines the amount of energy lost due to ionization \cite{elecionization}. $a$, $b$, and $\gamma$ are material-dependent constants. In air, $a = 147.55$, $b = 72.13$, $\gamma = 0.348$. $x$ and $X_0$ are the same as in Equation \ref{eqn:scatter}.
\begin{equation} \label{eqn:elecionization}
    -\frac{dE}{dx} = \left( \frac{m_e}{X_0} \right)\left( a - b \left(\frac{E}{m_e}\right)^{-\gamma}\right)
\end{equation}
Bremsstrahlung is governed by a stochastic process. When traveling through matter, charged particles can release a photon of energy $E_\gamma$, whose cross-section ($\frac{d\sigma}{dE_\gamma}$) is shown in Equation \ref{eqn:bremm} \cite{PPM}. In this cross-section, $N_A$ is Avogadro's number, $A$ is the weighted average of the atomic numbers of elements in the atmosphere, and $y$ is the ratio of $E_\gamma$ to the energy of the electron traveling through the atmosphere.   
\begin{equation} \label{eqn:bremm}
    \frac{d\sigma}{dE_\gamma} = \frac{A}{X_0 N_A E_\gamma} \left( \frac{4}{3}-\frac{4}{3}y+y^2 \right)
\end{equation}

\textbf{Cherenkov light production} -- Cherenkov light production for electrons follows the same process as for muons. 
    
\subsection{Analysis of Cherenkov Light Production}
\indent Each particle generates Cherenkov cones throughout its entire trajectory. The initial position, pointing direction, number of photons, and maximum Cherenkov angle ($\theta_{\text{max}}$) are all recorded at each step to reconstruct Cherenkov cones. The angular distribution for each cone is uniform in the range $(0.9 * \theta_{\text{max}}, \theta_{\text{max}}]$ to account for the fact that only the maximum angle was recorded. To analyze all Cherenkov cones together, we define a detection plane with origin at the detector and a normal vector along the optical axis. All cones are propagated until they reach this plane, where position on the plane, angle with respect to normal, and total time of flight (time since the primary entered the atmosphere) are recorded. To estimate the signal seen by a telescope on a balloon, we create a spatial bin on this plane and collect every photon that lands inside of it. We can then plot the angle and timing distributions of those photons.

To measure the magnetic field deflections of muons, we need a way to compare the muon component of the Cherenkov signal to the signal created by the main shower. The main shower is roughly estimated by creating Cherenkov cones along the optical axis with the number of Cherenkov photons determined by the GH distribution. The angular spread of these showers is over-estimated by using the $\beta$ value of a $10^{17}\: \text{eV}$ proton. This method allows us to compare the muon component of the shower to an approximation for the rest of the shower. In the future, we will implement a better-developed air shower simulation, such as EASCherSim \cite{Cummings_2021} to better estimate both the main shower Cherenkov signal, as well as the Cherenkov signals from the electrons. 

\section{Simulation Results \& Discussion}
\begin{figure}[h]
    \centering
    \includegraphics[width = \linewidth]{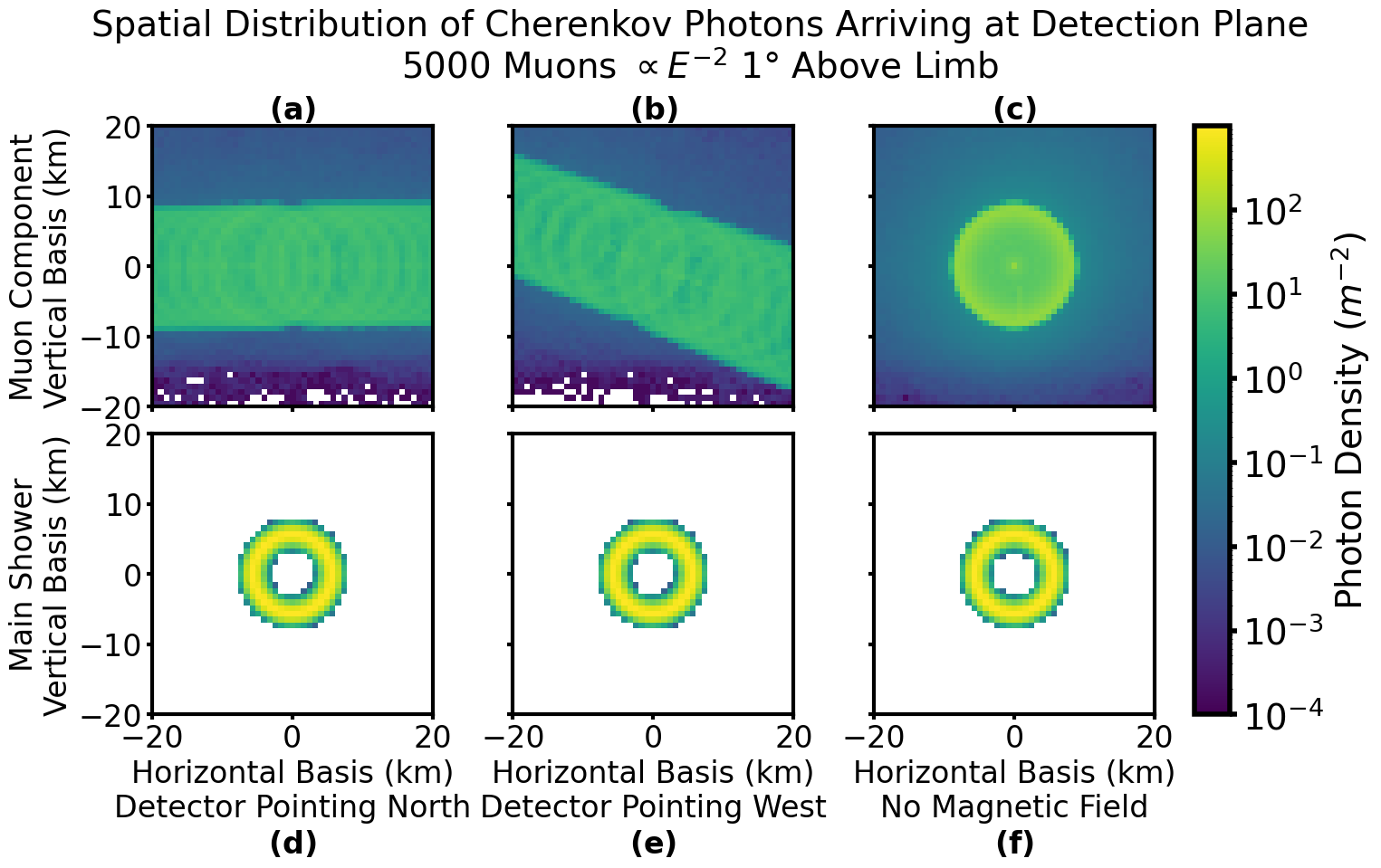}
    \caption{The simulated Cherenkov photons reaching a plane centered on the CT and perpendicular to the optical axis. (\textbf{a, b, c}) illustrate the Cherenkov spatial distributions for the muon component of the shower and (\textbf{d, e, f}) for the main shower Cherenkov estimate. (\textbf{a, b, d, e}) depict the two azimuthal pointing directions of the balloon and (\textbf{c, f}) show the control case with no magnetic field.}
    \label{fig:plane}
\end{figure}

\indent We chose the location of the detector in the simulation to be 33 km above sea level at the coordinates of Wanaka, New Zealand, the launch point of EUSO-SPB2. We simulated the detector pointing 1° above Earth's limb in zenith and both towards and perpendicular to magnetic North in azimuth. We also simulated a test case with no magnetic field as a sanity check. We chose to simulate 5,000 muons as a compromise between having enough statistics and the number of computational resources required. Using the power law described in Section 2.1, the total energy of muons summed up to about 400 TeV. For the main shower, we generated 50 billion Cherenkov photons as a rough estimate, which was about 5 times as many photons as the muon component created. This ratio between muons and the main shower Cherenkov signal is still being refined. A detection plane was created with origin at the detector location and a normal vector along the optical axis. Figure \ref{fig:plane} shows the simulated data plotted on this plane for each simulated configuration, including both the muon component and the estimated main shower component. At the location of the balloon, the magnetic field has strength $57 \:\mu \text{T}$ with declination 24$\degree$ and inclination -70\degree. The muon component of the Cherenkov spatial distribution indicates deflections left and right (depending on muon charge) when pointed North due to the radial component of Earth's magnetic field. Once rotated 90° in azimuth, the muon Cherenkov signal indicates an up/down deflection as well. With no magnetic field present, there are no deflections as expected.

\begin{figure}[h]
    \centering
    \includegraphics[width = 0.9\linewidth]{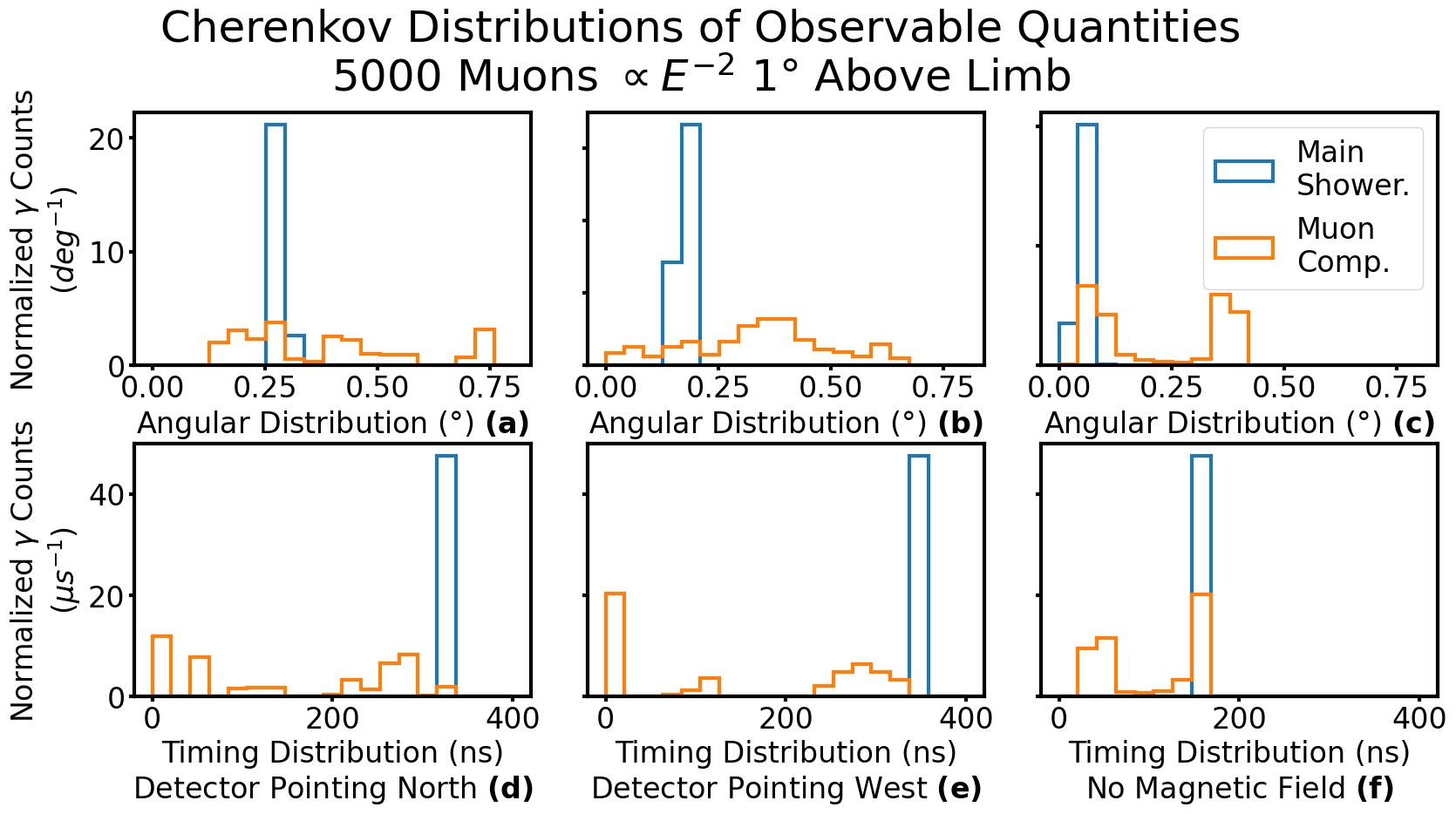}
    \caption{Angular (\textbf{a, b, c}) and Timing (\textbf{d, e, f}) distributions of Cherenkov light entering an arbitrarily-located 400 m by 400 m bin on the detection plane.}
    \label{fig:dist}
\end{figure}

To understand what a detector at a specific location would observe, we created a smaller 400 m x 400 m bin on the deflection plane and recorded every photon arriving inside of it. The bin location was arbitrary and was chosen in a place with a high density of photons from both components. The angular and timing distributions of those contained photons are displayed in Figure \ref{fig:dist}. 
These plots illustrate that there is a considerable difference between the Cherenkov timing distributions of the main shower and the muon component. The muon component of the Cherenkov signal arrives before the main component because the muons are traveling faster than the Cherenkov photons through the atmosphere. These timing distributions allow us to experimentally observe the \textit{muon tail} separate from the main shower, provided that the Cherenkov signal from the muons is above the optical background reaching the detector. The ratio between the intensity of the Cherenkov light from the muon component to the main shower is still unknown. We cannot conclude that we can experimentally measure the earlier Cherenkov signal from the \textit{muon tail} until we analyze other air shower simulations to determine if the Cherenkov signal from muons has high enough intensity to be above the background. 

The wider timing and angular spread for plots \textbf{(a, b, d,\& e)} compared to \textbf{(c \& f)} in Figure \ref{fig:dist} indicates that the geomagnetic field bends muons so they travel closer to the detector (at a faster speed than the photons) and their Cherenkov signal arrives at a wider variety of angles than with no magnetic field present. It is also important to note that these quantities are dependent on the location of the bin inside the Cherenkov cone, as well as the size of the detection bin. The CT onboard EUSO-SPB2 has a 1 $\text{m}^2$ aperture to collect light, an angular resolution of 0.4° per pixel, and a timing resolution of 10 ns \cite{ctpaper}. Our simulations currently cannot make conclusions on whether EUSO-SPB2 can measure deflections in the \textit{muon tail} because we do not have an adequate number of data simulated to scale our detection bin down to the size of the CT aperture. The CT also cannot control its location within any given Cherenkov cone, so we will need to sample multiple showers with bins at many locations to understand what effects can be seen by EUSO-SPB2.

\section{Conclusions \& Next Steps}
\indent Our simulation was able to model the angular and timing distributions of Cherenkov light from muons reaching a detector at 33 km altitude. The distributions identified attributes that could be used to experimentally observe the \textit{muon tail} and its magnetic field deflections. However, we cannot conclude whether EUSO-SPB2 can measure these effects until we fully understand the intensity of the muon Cherenkov signal and gather enough statistics to decrease our detection bin to reflect the aperture of the CT. Further improvements to the existing simulation framework will be made as well. The first of which will be using CORSIKA showers to create more realistic lateral momentum and energy distributions for the muons. We will also use EASCherSim \cite{Cummings_2021} to more accurately implement the Cherenkov signal generated by the primary shower, the electron after muon decay, and the correct ratio of the Cherenkov signal from muons to the main Cherenkov signal. Applying these adjustments will help us create results that reflect a more realistic model of the Cherenkov signal seen by EUSO-SPB2.\\

\noindent{\bf Acknowledgements} -- The authors acknowledge the support by NASA awards 11-APRA-0058, 16-APROBES16-0023, 17-APRA17-0066, NNX17AJ82G, NNX13AH54G, 80NSSC18K0246, 80NSSC18K0473,80NSSC19K0626, 80NSSC18K0464, 80NSSC22K1488, 80NSSC19K0627 and 80NSSC22K0426, the French space agency CNES, National Science Centre in Poland grant n. 2017/27/B/ST9/02162, and by ASI-INFN agreement n. 2021-8-HH.0 and its amendments. This research used resources of the US National Energy Research Scientific Computing Center (NERSC), the DOE Science User Facility operated under Contract No. DE-AC02-05CH11231. We acknowledge the NASA BPO and CSBF staffs for their extensive support. We also acknowledge the invaluable contributions of the administrative and technical staffs at our home institutions.

{\small
\bibliographystyle{JHEP-nt}
\bibliography{my-bib-database}}

\clearpage

\newpage
{\Large\bf Full Authors list: The JEM-EUSO Collaboration\\}

\begin{sloppypar}
{\small \noindent
S.~Abe$^{ff}$, 
J.H.~Adams Jr.$^{ld}$, 
D.~Allard$^{cb}$,
P.~Alldredge$^{ld}$,
R.~Aloisio$^{ep}$,
L.~Anchordoqui$^{le}$,
A.~Anzalone$^{ed,eh}$, 
E.~Arnone$^{ek,el}$,
M.~Bagheri$^{lh}$,
B.~Baret$^{cb}$
% ...
}
\end{sloppypar}
\vspace*{.3cm}

{ \footnotesize
\noindent
% ...
$^{cb}$ Universit\'e de Paris, CNRS, AstroParticule et Cosmologie, F-75013 Paris, France\\
% ...
$^{ed}$ Istituto Nazionale di Fisica Nucleare - Sezione di Catania, Italy\\
% ...
$^{eh}$ INAF - Istituto di Astrofisica Spaziale e Fisica Cosmica di Palermo, Italy\\
% ...
$^{ek}$ Istituto Nazionale di Fisica Nucleare - Sezione di Torino, Italy\\
$^{el}$ Dipartimento di Fisica, Universit\`a di Torino, Italy\\
% ...
$^{ep}$ Gran Sasso Science Institute, L'Aquila, Italy\\
% ...
$^{ff}$ Nihon University Chiyoda, Tokyo, Japan\\ 
% ...
$^{ld}$ University of Alabama in Huntsville, Huntsville, AL, USA\\
$^{le}$ Lehman College, City University of New York (CUNY), NY, USA\\
% ...
$^{lh}$ Georgia Institute of Technology, USA\\
% ...
}

\end{document}